\documentclass[preprint]{revtex4}
\usepackage{enumerate}
\usepackage{graphicx}
\usepackage{dcolumn}
\usepackage{bm}
\usepackage{amsmath}
\usepackage{amsxtra}
\usepackage{amstext}
\usepackage{amssymb}
\usepackage{latexsym}
\begin{document}

%%%%%%%%%%%%%%%%%%%%%%%%TITLE%%%%%%%%%%%%%%%%%%%%%%%%
\title{Rotation of Tokamak-Plasmas out of Mechanical Equilibria, in Absence of External Torques}

\author{
Giorgio SONNINO$^{1*}$, Alessandro CARDINALI$^{2}$, Alberto SONNINO$^{3}$ and Fulvio ZONCA$^{2}$\\
\vskip0.4truecm
$^{1}$Universit{\'e} Libre de Bruxelles (U.L.B.), Department of Theoretical Physics and Mathematics, Campus de la Plaine C.P. 231 - Bvd du Triomphe, 1050 Brussels - Belgium\\
\vskip0.1truecm
$^{2}$EURATOM-ENEA Fusion Association, Via E.Fermi 45, C.P. 65 - 00044 Frascati (Rome) - Italy\\
\vskip0.1truecm
$^{3}$Universit{\'e} Catholique de Louvain (UCL),  Ecole Polytechnique de Louvain (EPL), Rue Archim$\grave{\rm e}$de, 1 bte L6.11.01, 1348 Louvain-la-Neuve - Belgium}
%%%%%%%%%%%%%%%%%%%%%%END_TITLE%%%%%%%%%%%%%%%%%%%%%%

%%%%%%%%%%%%%%%%%%%%%%ABSTRACT%%%%%%%%%%%%%%%%%%%%%%%
\begin{abstract}
Rotation of tokamak-plasmas, not at the mechanical equilibrium, is investigated utilizing the Prigogine thermodynamic theorem. This theorem establishes that, for systems confined in rectangular boxes, the global motion of the system with barycentric velocity does not contribute to dissipation. This result, suitably applied to toroidally confined plasmas, suggests that the global barycentric rotations of the plasma, in the toroidal and poloidal directions, are pure reversible processes. In case of negligible viscosity and by supposing the validity of the balance equation for the internal forces, we show that the plasma, even not in the mechanical equilibrium, may freely rotate in the toroidal direction with an angular frequency, which may be higher than the neoclassical estimation. In addition, its toroidal rotation may cause the plasma to rotate globally in the poloidal direction at a speed faster than the expression found by the neoclassical theory. The eventual configuration is attained when the toroidal and poloidal angular frequencies reaches the values that minimize dissipation.

\vskip0.5truecm
\noindent *Email: gsonnino@ulb.ac.be

\end{abstract}

\maketitle
%%%%%%%%%%%%%%%%%%%%%END_ABSTRACT%%%%%%%%%%%%%%%%%%%%%

%%%%%%%%%%%%%%%%%%%%%%TEXT_PAPER%%%%%%%%%%%%%%%%%%%%%%

\section{Introduction}
\vskip 0.2truecm

Toroidal and poloidal plasma rotations are of great importance in fusion science as it plays a very important role for controlling transport in magnetically confined tokamak-plasmas \cite{rice}. Apart from improving the confinement, intrinsic plasma rotation is also important because it may affect the threshold to an improved confined state (L-H transition) \cite{terry} and, if combined with large rotational shear, it can produce the formation of an internal transport barriers (ITB) by suppression of turbulence \cite{hahm}. In addition, rotation is of significant interest for ITER performance. Recent experiments in the DIII-D tokamak have indeed shown that plasma rotation may suppress the resistive wall mode (RWM) in tokamaks \cite{strait}. The rotational velocity observed in tokamaks may be very high, especially in the toroidal direction, which may reach values of the speed of sound. Rotation in the poloidal direction is however much slower. In tokamaks, torque yielding rotation can have several origins, among which, heating systems such as Neutral Beam Injection (NBI) (see, for example, Ref.~\cite{nbi}) and radio-frequency heating (see, for example, Ref.~\cite{rfh}). However, even in absence of external torques, axisymmetric toroidal plasmas exhibit an intrinsic rotation both in the toroidal and the poloidal directions \cite{rozhansky}. Several mechanisms have been proposed for explaining the origin of the intrinsic plasma rotation. Without claim of completeness, we cite three of them. We first mention the theory where the turbulence driven residual stress, acting as an intrinsic torque, is supposed to spin-up toroidal rotation \cite{wang} (see also Ref.~\cite{kaye} for an overview). In this approach, the toroidal moment transport is driven by the parallel and the perpendicular Reynolds Stress, $\prec {\rm u}_r{\rm u}_\phi \succ$ (with ${\rm u}_r$ and ${\rm u}_\phi$ denoting the barycentric velocity in the radial and the toroidal directions, respectively, and $\prec\cdots\succ$ the {\it mean value operation}) \cite{diamond1}. According to this theory, the non-diffusive residual stress is responsible for the intrinsic plasma rotation, through a ${\rm k}_\parallel$-symmetry breaking mechanism able to convert radial inhomogeneity into parallel spectral asymmetry \cite{kwon}. Another approach can be found in Ref.~\cite{coppi}. This theory is based on the transfer of angular momentum to the wall and towards the centre of the plasma column by modes associated with the gradients of the plasma temperature and density. The inward transport of angular momentum from the edge of the plasma column to its centre is accounted for by the modes that are associated with the gradients of the plasma temperature and density in the presence of a velocity flow along the magnetic field. Finally, we cite the mechanism reported in Ref~\cite{callen}. In this theory, the resultant plasma toroidal rotation equation includes the effects of collision-induced perpendicular viscosities, anomalous transport due to micro-turbulence in the plasma and momentum sources. According to this picture, non-axisymmetric fields produce a toroidal torque throughout the plasma that relaxes the toroidal flow to an intrinsic ion-temperature-gradient diamagnetic-type flow in the direction counter to the plasma current. 

\noindent In this work, we do not address the validity of the (several) mechanisms proposed in literature for explaining the onset of the plasma spin-up. This paper treats (and solves) the following crucial questions on Tokamak-plasma rotation in absence of external torque.
\vskip 0.10truecm
\noindent {\it Even not in the mechanical equilibrium,

\noindent -  Why does the plasma may rotate without dissipations ?

\noindent - Why does the frequency in the toroidal direction may largely be different from the neoclassical estimation ? In this case, what is the role of the initial conditions ?

\noindent - Why, on the contrary, the value of the frequency in the poloidal direction tends to be damped towards the neoclassical value ?}
\vskip 0.10truecm

\noindent To the best of our knowledge, we cannot find in literature a rigorous answer to this series of questions. It should be noted that our results derive directly from rigorous thermodynamic theorems and they are not based upon a theory. In particular, our scope is to illustrate the Prigogine macroscopic thermodynamic theorem (PMT) which, in our opinion, may significantly contribute to the understanding of the features of the intrinsic rotation of toroidally confined plasmas. The PMT asserts that, for systems confined in rectangular boxes, the global motion of the system with barycentric velocity does not contribute to dissipation \cite{prigogineaes}, \cite{borkmans}. We shall adopt the single fluid picture and focus on the ion fluid, whose velocity can be considered equal to that of the single fluid, because the electron mass is negligible compared to the ion mass. By combining the results reported in Refs~\cite{balescu2} and \cite{sonnino} with the PMT, suitably readapted to tokamak-plasmas, we show that the barycentric rotation of dissipative tokamak-plasmas is a pure reversible process. In addition, if viscosity is negligible, we demonstrate that the global motion of the magnetically confined collisional plasmas, satisfying the balance equation for the internal forces, is determined up to an arbitrary flow velocity. In this respect, to avoid misunderstanding, we would like to clarify the following. It is known that an immediate application of the ideal magnetohydrodynamics to plasmas, {\it at the mechanical equilibrium}, leads to the result that the barycentric motion of the system is determined up to an arbitrary velocity. However, here we prove a stronger result. In absence of viscosity, the global motion of collisional plasmas is determined up to an arbitrary velocity, even if the system is {\it not} at the mechanical equilibrium. This theorem is a kind of thermodynamic equivalency version of the mechanical inertial principle, according to the scheme: ${\it Inertial\ Principle}\rightleftharpoons{\it Reversible\ Process}$. A direct consequence of this theorem, applied to tokamak-plasmas, leads to prove that the plasma may freely rotate with a toroidal angular frequency, which may be higher than the neoclassical prediction, with an arbitrary additional factor equal to $\Delta\omega_\phi\simeq{\rm v}_\phi/R_0$ (with ${\rm v}_\phi$ and $R_0$ denoting and arbitrary toroidal velocity and the major radius of the tokamak, respectively). In its turn, the toroidal rotation may cause the plasma to rotate globally in the poloidal direction at a speed faster than the expression found by the neoclassical theory, with an additional contribution equal to $\Delta\omega_\theta={\rm v}_\phi B_\theta/(rB_\phi)$ (where $B_\theta$ and $B_\phi$ stand respectively for the poloidal and the toroidal magnetic field configuration, and $r$ is the radial coordinate). Let us mention that the thermodynamic approach explains the Solomon experiment, without invoking the existence of an "anomalous torque" \cite{solomon}. Indeed, Solomon {\it et al.}, by studying the momentum confinement for the DIII-D tokamak as a function of applied neutral beam, found that, under balanced neutral beam injection ({\it i.e.} under zero total torque), the plasma still maintains a significant rotation in the co-direction. A preliminary work aiming to understand the plasma rotation effect through a thermodynamic approach, can be found in the footnote \footnote{G. Sonnino, {\it A Note on Plasma Rotation Based on the Thermodynamic Prigogine's Theorem}, submitted to publication in {\it Phys. Rev. Lett.} (2012).}.

\noindent The paper is organized as follows. In Section~\ref{thermo} we discuss the PMT and its main consequences. In particular, the Subsection \ref{thermoa} deals with dissipative systems, which are not in mechanical equilibrium. Conversely, the Subsection \ref{thermob} refers specifically to conservative diffusion processes occurring in systems at the mechanical equilibrium. Section~\ref{rotation}, is devoted to the application of the MPT to the intrinsic rotation effect in tokamak-plasmas. The frequency rotation in the poloidal direction is obtained in the Velikhov geometry \cite{velikhov} by removing any supplementary {\it ad hoc} hypothesis. Concluding remarks can be found in Section~\ref{cs}.

\vskip 0.2truecm
\section{Basic Equations for a Single Fluid Description and Prigogine's Theorem}\label{thermo}
\vskip 0.2truecm
We adopt the single fluid description, which excludes the issue of the momentum transfer between the ion and electron fluid. In order to set up the nomenclature, we briefly recall the basic definitions below. We assume that the barycentric velocity satisfies the evolution equation \cite{landau}
\begin{equation}\label{pr5}
\frac{d{\bf u}}{dt}={\bf a}\equiv-\rho^{-1}\nabla\cdot{\Pi}+{\bf\mathcal F}
\end{equation} 
\noindent with $\Pi$ denoting the pressure tensor measured by an observer moving with the {\it barycentric velocity} ${\bf u}$ \footnote{According to the demonstration of G{\'e}h{\'e}niau, for a system made by several species, the values of the scalar pressure and the pressure tensor depend on the definition of the average velocity \cite{geheniau}. In Eq.~(\ref{pr5}), $\Pi$ is the symmetric pressure tensor obtained by choosing the barycentric velocity as the average velocity.}. Also the substantial time-derivative, $d_t$, is constructed with the barycentric velocity :
\begin{equation}\label{pr6}
\frac{d}{dt}\equiv\frac{\partial}{\partial t}+{\bf u}\cdot\nabla
\end{equation}
\noindent The single fluid quantities, such as the {\it barycentric velocity}, {\bf u}, the {\it mass density}, $\rho$, the {\it barycentric force}, $\mathcal F$, and the ({\it density of\ }) {\it flux diffusion of species $\alpha$}, ${\bf J}^d_\alpha$ are defined in the usual way 
\begin{align}\label{pr4}
&{\bf u}\equiv\frac{m_i{\bf u}_i+m_e{\bf u}_e}{m_i+m_e}\quad;\quad
\rho=\sum_{\alpha=e,i}\rho_\alpha=\sum_{\alpha=e,i} n_\alpha m_\alpha\quad; \quad {\bf\mathcal F}=\sum_{\alpha=e,i}\frac{\rho_\alpha}{\rho}{\bf\mathcal F}_\alpha\\
&{\bf J}^d_\alpha=\rho_\alpha({\bf u}_\alpha-{\bf u})\qquad;\quad\ \!\alpha=(e,i)\qquad \nonumber
\end{align}
\noindent Here, $m_i$ and $m_e$, and ${\bf u}_i$ and ${\bf u}_e$ are the ion and the electron mass, and  the ion and the electron velocities, respectively. $q_e$ denotes the {\it absolute value of the electron charge} and $n_\alpha$ is the {\it species number density}. For the quasi neutrality condition, we have $n_e=Zn_i\simeq Zn$, with $Z$ denoting the charge number.
\vskip 0.2truecm
\subsection{Dissipative systems not in mechanical equilibrium}\label{thermoa}
\vskip 0.2truecm

\noindent The PMT holds by assuming the validity of the following hypothesis :
\vskip 0.2truecm
\noindent
{\it The Gibbs relation, valid for closed and open systems, also applies to the barycentric motion of the single fluid, under the condition that the derivatives appearing in this equation are modified according to the motion of the fluid i.e.},
\vskip 0.2truecm
\begin{equation}\label{pr7}
T\frac{ds}{dt}=\frac{du}{dt}+p\frac{dv}{dt}-\sum_{\alpha=e,i}\mu_\alpha\frac{dN_\alpha}{dt}
\end{equation}
\noindent {\it with the total derivatives provided by formula}~(\ref{pr6}). 

\noindent $s$ and $\mu_\alpha$ indicate the total entropy of the system per unit mass and the chemical potential (for species $\alpha$) per unit mass, respectively. $T$, $u$, $v$ and $N_\alpha$ are the temperature, the energy density per unit mass, the specific volume ($v=\rho^{-1}$) and the mass fractions, respectively. Prigogine demonstrated that, under the validity of this hypothesis, the entropy production $\sigma$, related to a continuos media, can be brought into the form \cite{prigogineaes}
\begin{align}\label{pr8}
&\sigma=\sum_{j=1,2,3}{\bf J}^h_j\frac{\partial}{\partial x_j}\Bigl(\frac{1}{T}\Bigr)-\frac{1}{T}\sum_{j\kappa=1,2,3}\pi_{j\kappa}\frac{\partial {\rm u}^\kappa}{\partial x_j}+\frac{1}{T}\sum_{\alpha=e,i}\sum_{j=1,2,3}\Bigl[{\mathcal F}^j_\alpha-\Bigl(\frac{\partial\mu_\alpha}{\partial x_j}\Bigr)_T\Bigr]J_{\alpha j}^d-\sigma_d\ >0\nonumber\\
&\sigma_d\equiv\frac{1}{T}\sum_{\alpha=e,i}\sum_{j=1,2,3}\Bigr[{\mathcal F}_\alpha^j-\Bigl(\frac{\partial\mu_\alpha}{\partial x_j}\Bigr)_T-{\mathcal R}_\alpha^j\Bigr]J_{\alpha j}^d
\end{align}
\noindent Here, indexes $j,\kappa$ distinguish the components of a vector or of the viscous tensor $\pi$. Moreover ($x_1,x_2,x_3$) stands for ($x,y,z$) [i.e., the three spacial cartesian coordinates]. ${\mathcal{\bf R}}_\alpha$ is the total force (with components $R_\alpha^j$) exerted on the particle species in question as a result of collision with other species in the plasma and ${\rm u}^\kappa$ denotes the components of the barycentric velocity. Hence, we have ${\mathcal{\bf R}}_\alpha=\sum_{\beta}{\mathcal{\bf R}}_{\alpha\beta}$ where ${\mathcal{\bf R}}_{\alpha\beta}$ represents the interaction between species $\alpha$ and $\beta$. ${\bf J}^h$ denotes the {\it entaltpy-heat flow linked to diffusion} defined as 
\begin{equation}\label{pr9}
{\bf J}^h\equiv {\bf J}^q+\sum_{\alpha=e,i}h_\alpha{\bf J}_\alpha^d
\end{equation}
\noindent with $h_\alpha$ denoting the enthalpy per unit mass of species $\alpha$, ${\bf J}^q$ the heat flow, and the flux diffusion of species $\alpha$, ${\bf J}_\alpha^d$, is defined in Eqs~(\ref{pr4}), respectively. The first term on the right hand side of the first equation of Eqs~(\ref{pr8}) expresses the dissipation due to the heat flow linked to diffusion, the second term represents the dissipation by viscosity of the barycentric kinetic energy, and the third term takes into account the dissipation by diffusion of the kinetic energy. Finally, the last contribution, $T\sigma_d$ in Eq.~(\ref{pr8}), is the source of the kinetic energy due to diffusion. Notice that the barycentric velocity does not appear explicitly in the entropy source strength Eq.~(\ref{pr8}). Expression (\ref{pr8}) contains only the derivatives $\partial_{x_i}{\rm u}^j$ and the diffusion vectors ${\bf J}_\alpha^d$ \footnote{Notice that, by definition, the diffusion vectors ${\bf J}_\alpha^d$ are invariant under transformation ${\bf u^i}\rightarrow {\bf u^i}+{\bf u}$.}. In other words, the global motion of the system with barycentric velocity ${\bf u}$ {\it does not} contribute to the entropy production. Hence, {\it the global motion of the system constitutes an "intrinsic" reversible processes}. This is the essence of the PMT. This theorem reads as follows. "{\it For systems confined in a rectangular box, global motion of the system with barycentric velocity ${\bf u}$ does not contribute to dissipation}" \cite{prigogineaes}, \cite{borkmans}. 

\noindent In neoclassical theory, plasma is considered to be in Onsager's region ({\it i.e.}, the closure relations are linear or, in other words, the thermodynamic flows are linearly connected with the thermodynamic forces) \cite{balescu}. According to the Minimum Entropy Production theorem \cite{prigoginemep}, in this region, for time-independent boundary conditions, the {\it state of minimum entropy production} turns out to be a {\it stationary state}. Because of Eq.~(\ref{pr8}) combined with the minimum entropy production theorem, also the steady-state reached by the system does not depend on the barycentric velocity (but only on its derivatives). Hence, with the addition of the Minimum Entropy Production Theorem, the PMT yields that in the Onsager region, the steady-state, obtained by minimizing the entropy production, remains unchanged by performing a global motion of the system with barycentric velocity. 

\noindent We immediately observe that in absence of viscosity and, for collisional plasmas where the balance equation for the internal forces is satisfied
\begin{equation}\label{prn}
\sum_{\alpha=e,i}\rho_\alpha {\mathcal R}_\alpha^j=0
\end{equation}
\noindent the entropy production of the system, Eq.~(\ref{pr8}), is invariant under the barycentric velocity transformation: ${\bf u}\rightarrow {\bf u}+{\bf v}$, with ${\bf v}$ denoting an arbitrary velocity. Hence for collisional plasmas, with no viscosity, global motion of the system is determined up to an arbitrary velocity ${\bf v}$. Notice that it is not required the condition that the system is at the mechanical equilibrium.

\vskip 0.2truecm
\subsection{Conservative diffusion processes occurring in systems at the mechanical equilibrium}\label{thermob}
\vskip 0.2truecm

\noindent In this subsection we shall consider the particular case of conservative diffusion processes occurring in systems at the mechanical equilibrium. We briefly check that, in these conditions and in case of no viscosity, the balance equation for the internal forces, Eq.~(\ref{prn}), is automatically satisfied. The source of the kinetic energy linked to diffusion, $T\sigma_d$, consists of three contributions. The first term takes into account the work made by the external forces and the second term is the contribution due to the gradients of concentration. Finally, the last term represents the dissipation of the diffusion kinetic energy due to the collisions between particles with different velocities. In many physical examples, the latter term compensates exactly the first two contributions {\it i.e.}, 
\begin{equation}\label{pr10}
{\mathcal R}_\alpha^j={\mathcal F}_\alpha^j-\Bigl(\frac{\partial\mu_\alpha}{\partial x_j}\Bigr)_T
\end{equation}
\noindent Consequently, $\sigma_d=0$. In literature, when condition (\ref{pr10}) is satisfied, the process is referred to as {\it diffusion conservative process}. Notice that, in case of diffusion conservative process, the entropy production given in Eq.~(\ref{pr8}) takes the form usually reported in literature (see, for example, Refs~\cite{borkmans} and \cite{glansdorff})
\begin{equation}\label{pr11}
\sigma=\sum_{j=1,2,3}{\bf J}^h_j\frac{\partial}{\partial x_j}\Bigl(\frac{1}{T}\Bigr)-\frac{1}{T}\sum_{j\kappa=1,2,3}\pi_{j\kappa}\frac{\partial {\rm u}^\kappa}{\partial x_j}+\frac{1}{T}\sum_{\alpha=e,i}\sum_{j=1,2,3}\Bigl[{\mathcal F}^j_\alpha-\Bigl(\frac{\partial\mu_\alpha}{\partial x_j}\Bigr)_T\Bigr]J_{\alpha j}^d\ >0
\end{equation}
\noindent When the barycentric acceleration {\bf a}, in Eq.~(\ref{pr5}) vanishes, the system is at the mechanical equilibrium. In absence of viscosity, the mechanical equilibrium condition reads
\begin{equation}\label{pr12}
\sum_{\alpha=e,i}\rho_\alpha {\bf \mathcal F}_\alpha^j=\frac{\partial p}{\partial x_j}
\end{equation}
\noindent However, from the Gibbs-Duhem relation
\begin{equation}\label{pr13}
\sum_{\alpha=e,i}\rho_\alpha d\mu_\alpha=-\rho s dT +d p
\end{equation}
\noindent we immediately obtain
\begin{equation}\label{pr14}
\sum_{\alpha=e,i}\rho_\alpha \Bigl(\frac{\partial\mu_\alpha}{\partial x_j}\Bigr)_T=\frac{\partial p}{\partial x_j}
\end{equation}
\noindent Hence, the mechanical equilibrium condition may be rewritten as 
\begin{equation}\label{pr15}
\sum_{\alpha=e,i}\rho_\alpha\Bigl[{\mathcal F}^j_\alpha-\Bigl(\frac{\partial\mu_\alpha}{\partial x_j}\Bigr)_T\Bigr]=0
\end{equation}
\noindent If the diffusion process is conservative, the mechanical equilibrium condition simply reads
\begin{equation}\label{pr16}
\sum_{\alpha=e,i}\rho_\alpha {\mathcal R}_\alpha^j=0
\end{equation}
\noindent where Eq.~(\ref{pr10}) has been taken into account. Hence, under the above {\it strict restrictions}, we re-obtain, as a very particular case, the result of the ideal magnetohydrodynamics applied to plasmas. At the mechanical equilibrium, global motion of a viscousless fluid, is determined up to an arbitrary velocity {\bf v}. Notice that in this particular case {\bf v} should satisfy the supplementary condition $d_t{\bf v}=0$ and its value is determined by specifying the initial conditions.
\vskip 0.2truecm
\section{Plasma Rotation}\label{rotation}
\vskip 0.2truecm
In this session we shall apply the results discussed in Section~\ref{thermo} to magnetically confined plasmas. First of all, we should check that the PMT, which has been proved for systems confined in a rectangular box, applies also to plasmas confined in toroidal geometry. 

\noindent $\bullet$ {\bf Entropy Production of L-Mode Tokamak-Plasmas}

\noindent Let us consider, for simplicity, fully ionized tokamak-plasmas, defined as a collection of magnetically confined electrons and positively charged ions, with the magnetic configuration written in the form \cite{balescu2}
\begin{equation}\label{e1}
{\bf B}=-\frac{1}{2\pi}\nabla\psi\times\nabla\phi+F(\psi)\nabla\phi
\end{equation}
\noindent Here $\psi$ indicates the {\it poloidal magnetic flux}, $F$ is the {\it characteristic of axisymmetric toroidal field} depending on the {\it surface function}, $\psi$, and $\phi$ is the poloidal angle, respectively. A fundamental remark is the fact that, even in a long mean free path regime, the (density of) distribution functions $f^\alpha({\bf v},{\bf x},t)$ of a magnetically confined plasmas are dominated by a local equilibrium reference state and, in the phase space $\{{\bf v},{\bf x}\}$, it depend on velocity only on through the diffusion vectors ${\bf J}^d_\alpha$ (see, for example, Ref.~\cite{balescu})
\begin{align}\label{f1}
&f^\alpha({\bf v},{\bf x},t)=f^{\alpha 0}({\bf J}^d_\alpha,{\bf x},t)[1+\chi^\alpha({\bf J}^d_\alpha,{\bf x},t)]\qquad\quad{\rm with}\nonumber\\
&f^{\alpha 0}({\bf J}^d_\alpha,{\bf x},t)=n_\alpha({\bf x},t)\Bigl(\frac{m_\alpha}{2\pi T_\alpha({\bf x},t)}\Bigr)^{3/2}\exp-\Bigl[\frac{1}{m_\alpha n^2_\alpha({\bf x},t) T_\alpha({\bf x},t)}\mid{\bf J}^d_\alpha\mid^2\Bigr]
\end{align}
\noindent The deviations from this state, $\chi^\alpha$ , can be expanded in a series of irreducible tensorial Hermite polynomials, $H_{\iota_1\cdots \iota_q}^{(m)}$, getting
\begin{align}\label{f2}
\chi^\alpha({\bf J}^d_\alpha,{\bf x},t)=&\sum_{\kappa=0}^\infty q^{\alpha(2\kappa)}({\bf x},t)H^{(2\kappa)}({\bf J}^d_\alpha)+\sum_{\kappa=0}^\infty q_{\iota_1}^{\alpha(2\kappa+1)}({\bf x},t)H_{\iota_1}^{(2\kappa+1)}({\bf J}^d_\alpha)\\
&+\sum_{\kappa=1}^\infty q_{\iota_1\iota_2}^{\alpha(2\kappa)}({\bf x},t)H_{\iota_1\iota_2}^{(2\kappa)}({\bf J}_\alpha^d)+\cdots\qquad\qquad ;\quad{\alpha}=(e,i)\nonumber
\end{align}
\noindent with $q^{\alpha(0)}=q^{\alpha(2)}=q_\iota^{\alpha(1)}\equiv0$ and summation over repeated tensor indices, $\iota_1,\iota_2,\cdots$, is understood. The dimensionless entropy production of species $\alpha$, $\sigma^\alpha$, is derived under the sole assumption that the state of the quiescent plasma is not too far from the reference local Maxwellian. In the {\it local dynamical triad} \cite{hinton3}, and up to the second order of the drift parameter $\epsilon=\rho_L/L\ll 1$ [with $\rho_L$ and $L\equiv\nabla^{-1}$ denoting respectively the Larmor radius and the length scale of variation of macroscopic plasma parameter (such as temperature, density, etc.)], $\sigma^\alpha$ can be brought into the form (see Refs~\cite{balescu2} and \cite{sonnino})
\begin{align}\label{e3}
\sigma^e=&\ q_{\parallel ps}^{(1)}(g_{\parallel}^{(1)}-{\bar g}_{\parallel}^{e(1)})+
q_{\parallel ps}^{e(3)}(g_{\parallel}^{e(3)}+{\bar g}_{\parallel}^{e(3)})
+q_{\parallel b}^{(1)}(g_{\parallel}^{(1)}-{\bar g}_{\parallel}^{e(1)})
+q_{\parallel b}^{e(3)}(g_{\parallel}^{e(3)}+{\bar g}_{\parallel}^{e(3)})\nonumber\\
&+q_{\parallel b}^{e(5)}{\bar g}_{\parallel}^{e(5)}+{\hat q}_{\rho cl}^{e(1)}g_{\rho}^{(1)P}+
{\hat q}_{\rho cl}^{e(3)}g_{\rho}^{e(3)}\\
\sigma^i=&\ q_{\parallel ps}^{i(3)}(g_{\parallel}^{i(3)}+{\bar g}_{\parallel}^{i(3)})\ +q_{\parallel b}^{i(3)}(g_{\parallel}^{i(3)}+{\bar g}_{\parallel}^{i(3)})+q_{\parallel b}^{i(5)}{\bar g}_{\parallel}^{i(5)}+{\hat q}_{\rho cl}^{i(3)}g_{\rho}^{i(3)}\nonumber
\end{align}
\noindent Here $q_\iota^{\alpha (\kappa)}$ [with the index $\iota$ taking the three possible directions: radial ($\rho$), tangent ($\parallel$) and per-tangent ($\wedge$)] denote the {\it Hermitian moments of the distribution functions} and $g_\iota^{\alpha (\kappa)},\ {\bar g}_\iota^{\alpha (\kappa)}$ are the {\it dimensionless source terms}. Index $\kappa$ takes the values $\kappa=(1,3,5)$. The source terms are related to the spatial inhomogeneity, and are {\it expressed as gradients of the fields} ({\it i.e.}, $\nabla n_\alpha$, $\nabla T_\alpha$, $\nabla {\bf u}_\alpha$ etc.). Hence, they depend on space and time (${\bf x},t$) only. We note that, the terms involving the radial thermodynamic forces are identical to those of the classical, thermodynamical form given by Eq.~(\ref{pr8}). Hence, in the drift approximation, the entropy production is obtained as the sum of two contributions: a {\it classical} piece, and a {\it neoclassical} piece. 
\begin{equation}\label{en3a}
\sigma^\alpha=\sigma_{cl}^\alpha+\sigma_{\parallel}^\alpha\qquad\qquad ;\quad\alpha=(e,i)
\end{equation}
\noindent In the (linear) Onsager region, Eqs~(\ref{e3}) reduce to \cite{balescu2}
\begin{align}\label{e4}
\sigma^e=&\ c_{13}^e(q_{\parallel}^{(1)})^2+c_{33}^e(q_{\parallel}^{e(3)})^2+c_{55}^e(q_{\parallel}^{e(5)})^2+2c_{13}^eq_{\parallel}^{(1)}q_{\parallel}^{e(3)}+2c_{15}^eq_{\parallel}^{(1)}q_{\parallel}^{e(5)}+2c_{35}^eq_{\parallel}^{e(3)}q_{\parallel}^{e(5)}
\nonumber\\
&+{\tilde\sigma}_\perp(g_\rho^{(1)P})^2+{\tilde\kappa}^e_\perp(g_\rho^{e(3)})^2
-2{\tilde\alpha}_\perp g_\rho^{(1)P}g_\rho^{e(3)}\nonumber\\
\sigma^i=&\ c_{33}^i(q_{\parallel}^{i(3)})^2+c_{35}^i(q_{\parallel}^{i(5)})^2+2c_{35}^eq_{\parallel}^{i(3)}q_{\parallel}^{i(5)}+{\tilde\kappa}^i_\perp(g_\rho^{i(3)})^2
\end{align}
\noindent with $c_{j\kappa}^\alpha$ denoting the collision matrix elements. Coefficients ${\tilde{\sigma}}_r$, ${\tilde{\alpha}}_r$, ${\tilde{\kappa}}^{\alpha}_r$ indicate the dimensionless component of the {\it electronic conductivity}, the {\it thermoelectric coefficient} and the {\it electric} ($\alpha =e$) or {\it ion} ($\alpha =i$) {\it thermal conductivity}, respectively. From Eqs~(\ref{f1}), (\ref{f2}) and (\ref{en3a}), we get
\begin{equation}\label{e5}
\sigma^\alpha=\sigma_{cl}^\alpha({\bf J}_\alpha^d,{\bf x},t)+\sigma_{\parallel}^\alpha({\bf J}_\alpha^d,{\bf x},t)\qquad\qquad ;\quad\alpha=(e,i)
\end{equation}
\noindent Hence, also for collisional tokamak-plasma, the barycentric velocity does not appear explicitly in the expression of the entropy production (see the footnote [42]). If the plasma becomes unstable and turbulent, this description becomes incomplete and supplementary contributions to the expression of the entropy production should be added \cite{pometescu}, \cite{kosuga}. However, we draw the attention to the fact that, also in the turbulent theory proposed in Ref.~\cite{diamond1}, the term which is supposed to be responsible of the intrinsic rotation is not linked to the velocity flow but it depends only on its derivatives. In the work of Y. Kosuga {\it et al.}, the term of the entropy source strength, ${\tilde\sigma}^T$, which is supposed to be responsible of the intrinsic plasma rotation can be brought into the bilinear ("{\it flux}$\ \times\ \!${\it force}") form : 
\begin{equation}\label{e5a}
{\tilde\sigma}^T=-\frac{n}{v_{thi}^2}\prec{\rm u}_r{\rm u}_\parallel\succ\frac{\partial \prec {\rm u}_\parallel\succ}{\partial r} 
\end{equation}
\noindent Here, $r$ is the radial coordinate of the tokamak. In the mean field approximation we have \cite{diamond1}
\begin{equation}\label{e5b}
\prec{\rm u}_r{\rm u}_\parallel \succ=-\chi_\phi \frac{\partial \prec {\rm u}_\parallel\succ}{\partial r} +\Pi_{r\parallel}^{res}
\end{equation}
\noindent Here, $\chi_\phi$ represent the {\it flow diffusivity} and $\Pi_{r,\phi}^{resid.}$ the {\it residual stress} piece, respectively. This latter contribution is not directly proportional to the toroidal rotation $ {\rm u}_\phi$ or its radial gradient, but it is driven by the gradients of temperature, pressure and number density. Hence 
\begin{equation}\label{e5c}
{\tilde\sigma}^T=\frac{n}{v_{thi}^2}\Bigl[\chi_\phi\Bigr(\frac{\partial \prec {\rm u}_\parallel\succ}{\partial r} \Bigl)^2- \Pi_{r\parallel}^{res}\frac{\partial \prec {\rm u}_\parallel\succ}{\partial r} \Bigr]
\end{equation}
\noindent According to the Diamond {\it et al.} theory, the contribution 
\begin{equation}\label{e5d}
{\tilde\sigma}^T_{Rot.}=-\frac{n}{v_{thi}^2}\Pi_{r\parallel}^{res}\frac{\partial \prec {\rm u}_\parallel\succ}{\partial r} \ <0
\end{equation}
\noindent is a negative definite quantity and it is supposed to be responsible of the onset of the intrinsic "spin-up" of the plasma \cite{kosuga}.

\noindent Our analysis leads then to the following conclusion: {\it The global motion of tokamak-plasmas with barycentric velocity constitutes an ÓintrinsicÓ reversible processes. The thermodynamic theorems discussed in the} Subsections~\ref{thermoa} {\it and} \ref{thermob} {\it also apply to plasmas, magnetically confined in toroidal geometry}.

\noindent $\bullet$ {\bf Onset of the "Spin-up" of the Plasma from Rest} 

\noindent We have shown that, if the Prigogine's hypothesis (see the Subsection~\ref{thermoa}) is fulfilled (and certain conditions satisfied), the results reported in Section~\ref{thermo} establish that a system may {\it either} stay at rest {\it or} undergo a global barycentric rigid motion without dissipation. Indeed, both situations are equally acceptable from the dynamical and thermodynamical point of view, because they satisfy Eq.~(\ref{pr5}) and they perform a reversible process ({\it i.e.}, they are not accompanied by dissipation). It will be the initial condition to select one between these two possibilities. However, many effects can drive a stationary plasma to rotate. The system may be subject, for example, to turbulence that can produce local Raynolds stress. The residual stress, driven by a certain radial inhomogeneity, leads to the formation of the intrinsic rotation working in concert with the boundary conditions \cite{diamond2}. Or, the excitation of modes driven by the plasma pressure gradient, may generate the inflow of angular momentum, whose source is at the edge of the plasma column \cite{coppi}. Ultimately, internal hydrodynamic fluctuations, with high value of the wave number ${\bf k}$, may also be enhanced by the nonlinear terms present in the balance equations and by the toroidal geometry \cite{landau}, \cite{malek}. Although, all of these mechanisms may explain the onset of the spin-up of the plasma from rest, we still have to understand the reason for which the plasma does not damp its rotation subsequently. Our task is, then, to estimate the solution of Eq.~(\ref{pr5}) with non vanishing initial flow velocity. Due to the predominance of the ion mass with respect to the electron mass, the velocity and momentum of the single fluid can be considered equal to the velocity and momentum of the ion fluid. Hence, we should solve the following evolution equation for the ion velocity 
\begin{equation}\label{prt1}
m_in_i\frac{d{\bf u}_i}{dt}= -\nabla p_i-\nabla\cdot\pi_i+nZq_e\Bigl({\bf E}+\frac{1}{c}{\bf u_i}\times{\bf B}\Bigr)+{\bf R}_e+{\bf f}_{ext}
\end{equation}
\noindent with initial condition ${\bf u}_{0i}\neq 0$. Here, $c$, ${\bf B}$ and ${\bf E}$ denote the speed of light, the {\it magnetic}, and the {\it electric} fields, respectively. ${\bf R}_e$ and ${\bf f}_{ext}$ are respectively the interaction with the electron fluid and the {\it external forces, per unit volume} (other than electromagnetic). Since our aim is to study the intrinsic plasma rotation, we set ${\bf f}_{ext}=0$. The scalar pressure $p$ is defined through the equation of state $p_i=n_iT_i\simeq nT$ and ${\bf\pi}_i$ denotes the ion viscous stress tensor. 

\noindent $\bullet$ {\bf Toroidal Rotation} 

\noindent Our aim is to determine the toroidal angular frequency by applying the result shown at the end of the Subsection \ref{thermoa} and the Subsection \ref{thermob}. In absence of external forces and by imposing the validity of the balance equation for the internal forces, the perpendicular velocity ${\bf u}_\perp$ is immediately obtained from Eq.~(\ref{prt1}) by taking the vectorial product with {\bf B} (see also the footnote \footnote{Notice that from the Subsection~\ref{thermob}, a system satisfying the balance equation for the internal forces and where a diffusion conservative process occurs, is at the mechanical equilibrium.}). At the leading order of the drift parameter, the perpendicular component of the flow equals the ${\bf E}\times{\bf B}$ drift and the diamagnetic flow (see, for example, Refs \cite{sigmar} and \cite{hua})
\begin{equation}\label{prt2}
{\bf u}_\perp\simeq {\bf u}_{\perp i}\simeq \frac{{\bf E}\times{\bf b}}{B}+\frac{{\bf b}\times \nabla P}{m_in\Omega_i}\qquad{\rm with}\quad {\bf b}\equiv \frac{{\bf B}}{B}
\end{equation}
\noindent Here $\Omega_i$ denotes the ion gyrofrequency. Hence, at the lowest order of $\epsilon$ and {\it in the absence of poloidal flow}, the toroidal rotation frequency, $\omega_{\phi\ \!\!n.c.}$, is 
\begin{equation}\label{prt3}
\omega_{\phi\ \!\!n.c.}= -\frac{d\Phi}{d\psi}-\frac{1}{nq_e}\frac{dp}{d\psi}
\end{equation}
\noindent where $\Phi$ and $\psi$ denote the electrostatic potential and the poloidal flux, respectively. Hence, a driven toroidal plasma rotation is established in plasmas due to the formation of a radial electric field (in the presence of an externally applied magnetic field) and to a pressure gradient in the plasma. The plasma may then freely rotate in the toroidal direction, with frequency approximatively equal to 
\begin{equation}\label{prt3a}
\omega_\phi=\omega_{\phi\ \!\!n.c.}+{\rm v}_\phi/R
\end{equation}
\noindent with {\bf v} denoting an {\it arbitrary} "departure" of the ion velocity from the barycentric velocity ${\bf u}$, and $R=R_0+r\cos\theta$ ($\theta$ is the poloidal angle). In other words, plasma can freely rotate with an angular toroidal frequency, which may be increased with an arbitrary contribution $\Delta\omega_\phi={\rm v}_\phi/R$ with respect to the neoclassical expression reported in Eq.~(\ref{prt3}). The initial conditions determine the value of {\bf v}. In this respect, we mention that experimental measurements of toroidal rotation have been conducted in the L-mode, TCV ("{\it tokamak {\`a} configuration variable}") -plasmas, without external momentum injections \cite{scarabosio}. It has been found that the neoclassical theory of toroidal rotation disagrees with the experimental observations qualitatively and quantitatively. It should be reminded, however, that if  viscosity is taken into account the plasma may rotate toroidally with angular velocity approximatively given by Eq.~(\ref{prt3}). 

\noindent $\bullet$ {\bf Poloidal Rotation} 

\noindent Let us now determine the poloidal angular frequency by applying the result discussed in the Subsection \ref{thermoa}. To determine the poloidal velocity flow, we should take into account the following two observations :

\noindent {\bf j)} In mechanical equilibrium, the barycentric velocity for plasma with no viscosity, inertial, nor external forces, obeys to the equation
\begin{equation}\label{prt4}
 \nabla p\simeq \nabla p_i=nZq_e\Bigl({\bf E}+\frac{1}{c}{\bf {\rm u}_i}\times{\bf B}\Bigr)
\end{equation}
\noindent By rewriting Eq.~(\ref{prt4}) in terms of the toroidal and the poloidal velocity components ($u_r,u_\phi,u_\theta$), we obtain the following relation between the angular frequencies
\begin{equation}\label{prt5}
\omega_\theta=\omega_\phi\frac{B_\theta}{B_\phi}\frac{R}{r}+\frac{E_r}{rB_\phi}-\frac{\nabla p\cdot{\bf e}_r}{nrZq_eB_\phi}
\end{equation}
\noindent with ${\bf e}_r$ denoting the unit vector in the radial direction. Eq.~(\ref{prt5}) relates the poloidal and toroidal rotation to each other. In general it is not possible for the plasma to rotate poloidally without also rotating toroidally.

\noindent {\bf jj)} In Eq.~(\ref{prt1}) we made use of the fact that all terms in the viscosity tensor are small (by a factor $\epsilon$) with respect to the other terms in the momentum equation, $\nabla\cdot\pi\ll\nabla p$. However, this does not means that viscosity may always be neglected. In tokamaks, the parallel pressure gradient vanishes at the leading order and parallel viscosity plays a crucial role in determining the parallel barycentric velocity. 

\noindent An estimation of the order of magnitude of the barycentric velocity in the poloidal direction may be easily obtained by analyzing plasmas in the two-dimensional Velikhov model \cite{velikhov}. The model is constructed by taking into account considerations {\bf j)} and {\bf jj)}. We consider a plane layer of plasma, which is in motion along the $x$-axis [which, according to {\bf j)}, simulates the presence of a "toroidal rotation"], but with no initial sliding motion in the $y$ direction (this direction simulates a "poloidal rotation"). We suppose that the layer is subject to a magnetic field directed along the $x$-axis. In this case, the ion velocity, in the y-direction, satisfies the following simple equation, where the viscosity pressor tensor is no longer neglected [according to {\bf jj)}]
\begin{equation}\label{prt6}
m_in_i\frac{d{\rm u}_{iy}}{dt}= -\frac{\partial}{\partial x}\pi_i^{yx}
\end{equation}
\noindent Eq.~(\ref{prt6}) reads
\begin{equation}\label{prt7}
m_in_i\Bigl(\frac{\partial }{\partial t}+{\rm u}_{ix}\frac{\partial}{\partial x}\Bigl){\rm u}_{iy}=
\frac{\partial}{\partial x}\eta_i^{(1)}\frac{\partial {\rm u}_{iy}}{\partial x}-\frac{\partial}{\partial x}\eta_i^{(3)}\frac{\partial {\rm u}_{ix}}{\partial x}
\end{equation}
\noindent The viscosity coefficients for ions are $\eta_i^{(1)}=3n_iT_i/(10\Omega_i^2\tau_i)$ and $\eta_i^{(3)}=n_iT_i/(2\Omega_i)$ (with $\tau_i$ denoting the ion-ion collision time) \cite{chapman}, \cite{sigmar}. In the low-collisional regime we have $\Omega_i\tau_i\gg1$ and $\eta_i^{(1)}$ can be neglected in this transport regime. Using this approximation, we obtain the following equation of continuity for the ion momentum in the y-direction
\begin{equation}\label{prt8}
\frac{\partial p_{iy}}{\partial t}+\frac{\partial}{\partial x}{\rm u}_{ix} p_{iy}=\Bigl(p_{iy}-\frac{1}{n_i}\frac{\partial}{\partial x}\frac{n_iT_i}{2\Omega_i}\Bigr)\frac{\partial {\rm u}_{i}}{\partial x}-\frac{T_i}{2\Omega_i}\frac{\partial^2 {\rm u}_{ix}}{\partial x^2}
\end{equation}
\noindent where $p_{iy}=m_iu_{iy}$. We should now introduce the definition of the barycentric momentum in the y-direction, $p_{yb}$. We have already mentioned that, for a system made by several species, the viscous tensor depends on the velocity of the reference frame \cite{geheniau}. In the single fluid description, the barycentric momentum is defined so as to satisfy the continuity equation where the transport coefficients of the viscous stress tensor are considered as constants and the compressibility of the fluid is negligible \cite{borkmans}. Somehow, this definition "simulates" a global motion of the system. The viscous stress tensor is then constructed with the derivatives of the departure of the ion velocity from this barycentric velocity. Hence, in our two-dimensional model, the dynamic equation for the barycentric momentum of the fluid in the y-direction reads
\begin{equation}\label{prt9}
n\Bigl(\frac{\partial p_{yb}}{\partial t}+\frac{\partial}{\partial x}{\rm u}_{xb} p_{yb}\Bigr)=-\eta^{(3)}\frac{\partial^2}{\partial x^2}{\rm u}_{xb}
\end{equation}
\noindent with $\eta^{(3)}=nT/(2\Omega_i)\simeq n_iT_i/(2\Omega_i)=\eta^{(3)}_i$. By imposing that the momentum of the single fluid is equal to the momentum of the ion fluid, we find that the system performs a "global" sliding motion ("rotation") for $n\simeq n_i$ and $u_{yb}=u_{iy}$, with $u_{yb}$ given by
\begin{equation}\label{prt10}
{\rm u}_{yb}=\frac{1}{n_i}\frac{\partial}{\partial x}\frac{n_iT_i}{2m_i\Omega_i}\simeq\frac{0.5}{m_i\Omega_i}\frac{\partial T_i}{\partial x}
\end{equation}
\noindent In toroidal coordinates ($r,\theta,\phi$), a more accurate calculation for tokamak-plasmas with circular cross section and large aspect ratio, gets (see, for example, \cite{sigmar} and \footnote{The numerical coefficient, estimated by the neoclassical theory, is $-0.5$ in the plateau regime and $-1.7$ in the fully collisional regime (see, for example, \cite{sigmar}).}) 
\begin{equation}\label{prt11}
{\rm u}_{\theta\ \!\!n.c.}=\frac{1.17}{m_i\Omega_i}\frac{d T_i}{d r}
\end{equation}
\noindent In order of magnitude, we obtain the following neoclassical expression for the poloidal angular frequency
\begin{equation}\label{prt12}
\omega_{\theta\ \!\!n.c.}\simeq\frac{1.17}{rm_i\Omega_i}\frac{d T_i}{dr}
\end{equation}
\noindent Notice that, with respect to the original work \cite{velikhov}, we have removed the {\it ad hoc} hypothesis stating that $T/\Omega_i$ is an adiabatic invariant. According to the results in the Subsection~\ref{thermoa}, {\it the plasma freely rotates poloidally, with angular velocity given by} Eq.~(\ref{prt12}), {\it without producing dissipation}. Expression (\ref{prt12}) roughly coincides with the experimental observations. However, since $u_y$ is the drift velocity of ions, according to Rosenbluth, a plasma moving with such velocity must be only {\it marginally stable} \cite{rosenbluth}. Taking into account Eq.~(\ref{prt3a}) and Eq.~(\ref{prt5}), the toroidal rotation will tend to increment the neoclassical expression for the poloidal rotation [{\it i.e.}, Eq.~(\ref{prt12})], with an additional contribution equal to
\begin{equation}\label{prt13}
\Delta\omega_{\theta}\simeq \frac{B_\theta}{B_\phi}\frac{{\rm v}_\phi}{r}
\end{equation}
\noindent Therefore, according to the equation for plasmas in mechanical equilibrium, the toroidal rotation may cause the plasma to rotate poloidally at a speed faster than the neoclassical expression given by Eq.~(\ref{prt12}). However, the result discussed in the Subsection~\ref{thermoa} tell us that the departure of the poloidal angular frequency from the neoclassical value affects the value of the entropy production by contributing to dissipation. In addition, collisions tend to restore the neoclassical value. Hence, the magnitude of the poloidal angular frequency is determined by the toroidal rotation (which drives the poloidal rotation) and by the neoclassical poloidal flow damping. Nevertheless, a variation of the poloidal angular rotation modifies, in its turn, the value of the toroidal angular frequency [through Eq.~(\ref{prt5})], and so on. At the end, this mutual interplay breaks off when the toroidal and poloidal angular frequencies attains the values that minimize dissipation.

\noindent
\vskip 0.2truecm
\section{Conclusions}\label{cs}
\vskip 0.2truecm 

\noindent From the results reported in Refs~\cite{balescu2}, \cite{sonnino}, and the Prigogine theorems, suitably readapted to toroidally confined dissipative plasmas, we showed that the global barycentric rotation does not contribute to dissipation. In case of negligible viscosity, we demonstrate that the global motion of tokamak-plasmas, satisfying the balance equation for the internal forces, is determined up to an arbitrary flow velocity. The validity of these results do not require that the system is at the mechanical equilibrium. In these conditions, the plasma may freely rotate in the toroidal direction with an angular frequency that may be higher than the neoclassical estimation, with an additional contribution equal to $\Delta\omega_\phi={\rm v}_\phi/R$ (with ${\rm v}_\phi$ denoting an {\it arbitrary} velocity). This effect may help to explain the origin of the observed discrepancy between the neoclassical predictions and the experimental measurements in L-mode TCV plasmas. Plasma may also undergo a self-generated rotation in the poloidal direction with the angular frequency found by the neoclassical theory. Also in this case, the phenomenon is not accompanied by dissipation. However, from the Rosenbluth theory, a plasma moving close to this drift velocity is only marginally stable. The toroidal rotation may, then, force the plasma to increment the poloidal angular frequency with an additional contribution equal to $\Delta\omega_\theta=\Delta\omega_\phi RB_\theta/(rB_\phi)$. The poloidal rotation, in its turn, will affect, via Eq.~(\ref{prt5}), the value of the toroidal angular frequency and so on. The eventual configuration will be attained when the toroidal and poloidal angular frequencies reaches the values that minimize dissipation. 

\begin{figure*}[htb]\resizebox{0.55\textwidth}{!}{%
\includegraphics{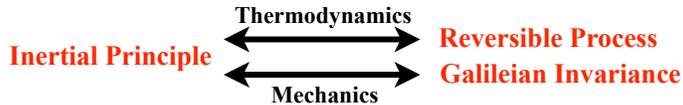}}
\caption{\label{rp} In thermodynamics, a {\it reversible process} is "equivalent" to the {\it Galileian invariance} valid in mechanics, in the sense that both express the concept of {\it inertial principle}.}
\end{figure*}

\noindent The next task should be, then, to determine the evolution equations for toroidal and poloidal angular frequencies and to analyze the stability of the mechanical equilibrium (or of the steady state solution). The study of this problem is beyond the scope of the present work. We have seen that, according to our description, two configurations of the plasma are compatible with the dynamical laws and thermodynamics. The plasma may {\it either} stay at rest {\it or} to perform a global motion at the barycentric velocity. Both configurations are allowed by the thermodynamical principles because they are both related to {\it reversible processes}. As mentioned in the introduction, one of the leading theory suggests that the residual stress may spin-up  the plasma from rest, acting in synergy with radial boundary conditions \cite{gurcan} (see also the footnote \footnote{It should be mentioned, however, that this mechanism, concerning the origin of the spin-up, is not in accordance with Velikhov's point of view \cite{velikhov}.}). However, even if the above mentioned theory may suggest a mechanism able to interpret the origin of the rotation of the the tokamak-plasma in absence of external torque, it remains an open question the reason for which the plasma is able to self-sustain a global motion with a well determined angular frequency. The latter question is addressed in this work. We tackled this intricate problem by means of Prigogine's thermodynamic theorem, and by identifying the collisions as the only source of irreversibility or dissipation. Prigogine's theorem suggests that, once established the onset of the spin-up of the plasma, the system can self-maintain a global motion at the barycentric velocity since this process is not accompanied by dissipation. This theorem is a kind of thermodynamic equivalency version of the (mechanical) Galilean inertial principle, according to the scheme illustrated in FIG.~\ref{rp}.
%%%%%%%%%%%%%%%%%%%%%%%%%%%%%%%%%%%%%%%%%%%%%%%%
%%%%%%%%%%%%%%%%%%%%%%%%%%%%%%%%%%%%%%%%%%%%%%%%
\noindent We have remarked that the turbulent contributions, estimated by the Diamond {\it et Al.} theory, do not affect the validity of the results reported in Section~\ref{thermo}. Moreover, the thermodynamic interpretation is perfectly in line with the experiment conducted on the DIII-D plasma, subject to external NBI \cite{solomon}. 

\noindent We conclude this section by mentioning that the paper cited in the footnote \footnote{V.G. Nosov and A.M. Kamchatnov (2008), {\it Thermodynamic inequalities for nuclear rotation}, http://arxiv.org/pdf/nucl-th/0408029.pdf.}, approaches the problem of obtaining a number of bounds of rotational levels in non spherical nuclei in a spirit similar to the one outlined above.
\section{Acknowledgments}
One of us (G.S.), is very grateful to Prof. M. Malek Mansour and Prof. R. Lefever, of the Universit{\'e} Libre de Bruxelles (U.L.B.), for scientific suggestions, and Dr U. Finzi, for encouraging discussions. We also thank Dr. M. Marinucci of the EURATOM/ENEA Fusion Association in Frascati (Rome) and Dr. G. Breyiannis for having deeply read the manuscript.

%%%%%%%%%%%%%%%%%%%%%END_TEXT_PAPER%%%%%%%%%%%%%%%%%%%%

%%%%%%%%%%%%%%%%%%%%%BIBLIOGRAPHY%%%%%%%%%%%%%%%%%%%%%%

%%%%%%%%%%%%%%%%%%%%%END_BIBLIOGRAPHY%%%%%%%%%%%%%%%%%%%


\begin{thebibliography}{alpha}

\bibitem{rice} J.E. Rice {\it et al.}, {\it Nuclear Fusion}, {\bf 47}, 1618 (2007).

\bibitem{terry} P.W. Terry {\it PRev. Mod. Phys.}, {\bf 72}, 109 (2000).

\bibitem{hahm} T.S. Hahm, {\it Phys. Plasmas}, {\bf 1}, 2940 (1994).

\bibitem{strait} E.J. Strait {\it et al.}, {\it Phys. Plasmas}, {\bf 14}, 056101 (2007).

\bibitem{nbi} F.L. Hinton and M.N. Rosenbluth, {\it Physics Letters A}, {\bf 259}, 267 (1999).

\bibitem{rfh} R. Cesario, A. Cardinali, C. Castaldo, M. Leigheb {\it et al}., {\it Phys. Plasmas}, {\bf 8}, 4721 (2001).

\noindent J.R. Myra and D.A. D'Ippolito, {Phys. Plasmas}, {\bf 7}, 3600 (2000).

\noindent J.R. Myra and D.A. D'Ippolito, {Phys. Plasmas}, {\bf 9}, 3867 (2002).

\bibitem{rozhansky} V. Rozhansky and M. Tendler, {\it Review of Plasma Physics}, {\bf 19}, 147 (1996).

\noindent {lin} Y. Lin {\it et al.}, {\it Phys. Plasmas}, {\bf 16}, 056102 (2009).

\bibitem{wang} W. X. Wang {\it et al.}, {Physics of Plasmas}, {\bf 18}, 042502 (2011).

\bibitem{kaye} S. Kaye (2012), {\it  Why Magnetically Confined Plasmas Rotate and Why it is Important}, Bulletin of the American Physical Society, {\bf 57}, Number 3, Atlanta, Georgia.

\bibitem{diamond1} P.H. Diamond {\it et al.}, {\it Nucl. Fusion}, {\bf 49}, 045002 (2009).

\bibitem{kwon} J.M. Kwon, S. Yi, T. Rhee, P.H. Diamond, K. Miki, T.S. Hahm, J.Y. Kim, {\"{O}}.D. G{\"{u}}rcan and C. McDevitt, {\it Nuclear Fusion}, {\bf 52}, 013004 (2012).

\bibitem{coppi} B. Coppi, {\it Nuclear Fusion}, {\bf 42}, 1 (2001). 

\bibitem{callen} J.D. Callen, A.J. Cole and C.C. Hegna, {\it Nuclear Fusion}, {\bf 49}, 08502 (2009).

\bibitem{prigoginemep} I. Prigogine, 1954 {\it Introduction to Thermodynamics of Irreversible processes}, (John Wiley \& Sons).

\bibitem{borkmans} C. Vidal, G. Dewel and P. Borckmans (1994), {\it Au-del{\`a} de l'{\'e}quilibre}, Hermann Editeurs des Sciences et des Arts, Paris. 

\bibitem{balescu2} R. Balescu, 1988 {\it Transport Processes in Plasmas. Vol 2. Neoclassical Transport}, Elsevier Science Publishers B.V., Amsterdam, North-Holland. 

\bibitem{sonnino} G. Sonnino and P. Peeters {\it Physics of Plasmas} {\bf 15}, 062309/1-062309/23 (2008).

\bibitem{solomon} W. M. Solomon {\it et al.}, {\it Plasma Phys. and Control. Fusion}, {\bf 49}, B313 (2007).

\bibitem{velikhov} E.P. Velikhov, {\it Atomnaya Energiya}, {\bf 14}, 573 (1963).

\bibitem{landau} L.D. Landau and E.M. Lifshitz (1987), {\it Fluid Mechanics. Course of Theoretical Physics; Vol. 6}, 2nd edition, Elsevier plc group, Oxford.

\bibitem{geheniau} J. G{\'e}h{\'e}niau, {\it Bull. Ac. Roy. Blg., Cl. Sc.}, Session of 11 April 1942 and of 9 january 1943.

\bibitem{prigogineaes} I. Prigogine (1947), {\it Etude Thermodynamique des Ph{\`e}nom{\`e}nes Irr{\'e}versibles}, Th{\`e}se d'Aggr{\'e}gation de l'Einseignement Sup{\'e}rieur de l'Universit{\'e} Libre de Bruxelles (U.L.B.).

\bibitem{balescu} R. Balescu, 1988 {\it Transport Processes in Plasmas. Vol 1. Classical Transport}, Elsevier Science Publishers B.V., Amsterdam, North-Holland. 

\bibitem{glansdorff} P. Glansdorff and I. Prigogine (1971), {\it Thermodynamic Theory of Structure, Stability and Fluctuations}, Wiley-Interscience, London-New york-Sydney-Toronto.

\bibitem{hinton3} F.L. Hinton and R.D. Hazeltine, {\it Rev. Mod. Phys.}, {\bf 48}, 239, (1976).

\bibitem{pometescu} N. Pometescu, {\it Plasma Phys. Control. Fusion}, {\bf 41}, 1453 (1998).

\bibitem{kosuga} Y. Kosuga, P.H. Diamond, and {\"{O}}.D. G{\"{u}}rcan, {\it Physics of Plasmas}, {\bf 17}, 102313 (2010).

\bibitem{diamond2} P.H. Diamond {\it et al.}, {\it Phys. Plasmas}, {\bf 15}, 012303, (2008).

\bibitem{malek} M. Malek Mansour (1991), {\it Probl{\`e}mes d'actualit{\'e} de la Phyisue II. Dynamique des fluides: aspects microscopiques et simulations}, Lectures notes, "{\it Deuxi{\`e}me Licence en Sciences Physiques}", Universit{\'e} Libre de Bruxelles (U.L.B.).

\bibitem{sigmar} P. Helander and D. Sigmar (2002), {\it Collisional Transport in Magnetized Plasmas}, Cambridge University Press,Cambridge, New York.

\bibitem{hua} M-D. Hua (2009), {\it Plasma Rotation in the MUST and JET Tokamaks}, Ph.D thesis, University of London and Diploma of Membership of the Imperial College.

\bibitem{scarabosio} A. Scarabosio, A. Bortolon, B.P. Duval, A. Karpushov and A. Pochelon, {\it Plasma Phys. Control. Fusion}, {\bf 48}, 663, (2006).

\bibitem{chapman} S. Chapman and T.J. Cowling (1961), {\it Mathematical Theory of Non-Uniform Gases}, Third edition, Cambridge University Press, Cambridge, New York.

\bibitem{rosenbluth} N. Krall, N. Rostoker and M. Rosenbluth, {\it Nuclear Fusion}, Suppl. Part 1, 143 (1962).

\bibitem{hinton} F.L. Hinton and S.K. Wong, {\it Phys. of Fluids}, {\bf 28}, 3082 (1985).

\bibitem{gurcan} O.D. Gurcan, P.H. Diamond, T.S. Hahm, and R. Singh, {\it Physics of Plasmas}, {\bf 14}, 042306, 2007 

\end{thebibliography}
\end{document}